\documentstyle[12pt]{article}
\let\section=\subsection     \let\subsection=\subsubsection                
\newcommand{\spc}{{\ }}

\newcommand{\pr}[1]{{\sc{\lowercase{#1}}}}

\newcommand{\Dy}{$^{152}$Dy}

\newcounter{leteq}
\newcommand{\steplet}{\stepcounter{leteq}\addtocounter{equation}{-1}}

\newenvironment{eqnalpha}{\setcounter{leteq}{1}

\begin{eqnarray}}{\end{eqnarray}%
}

\newenvironment{eqnalphalabel}[1]{\setcounter{leteq}{1}
\raisebox{0cm}[0cm][0cm]{\begin{minipage}{1cm}%
\begin{eqnarray}\label{#1}&&\nonumber\end{eqnarray}\end{minipage}}

\begin{eqnarray}}{\end{eqnarray}%
}

\newcommand{\bnl}{\begin{eqnalpha}}
\newcommand{\enl}{\end{eqnalpha}}
\newcommand{\bnll}[1]{\begin{eqnalphalabel}{#1}}
\newcommand{\enll}{\end{eqnalphalabel}}

\newcommand{\bbox}[1]{\mbox{{\boldmath{$#1$}}}}
\newcommand{\cbox}[1]{\mbox{{\scriptsize{\boldmath{$#1$}}}}}
\newcommand{\text}[1]{\mbox{\scriptsize{#1}}}
\newcommand{\tensor}[1]{\stackrel{\leftrightarrow}{#1}}

\newcommand{\be}{\begin{equation}}
\newcommand{\ee}{\end{equation}}
\newcommand{\ba}{\begin{array}}
\newcommand{\ea}{\end{array}}
\newcommand{\bn}{\begin{eqnarray}}
\newcommand{\en}{\end{eqnarray}}
\newcommand{\bc}{\begin{center}}
\newcommand{\ec}{\end{center}}
\newcommand{\bi}{\begin{itemize}}
\newcommand{\ei}{\end{itemize}}

\newcommand{\thalf}{{\textstyle{\frac{1}{2}}}}
\newcommand{\third}{{\textstyle{\frac{1}{3}}}}

\begin{document}

\hfill{CRN 96-30}

\vspace{0.5cm}
\begin{center}
        {\bf\Large
                     Solution of the Skyrme-Hartree-Fock equations in
                     the Cartesian deformed harmonic oscillator basis.
                                      (I) The method.
        }

\vspace{5mm}
        {\large
                       J.Dobaczewski$^{a,b,}$\footnote
                       {E-mail: dobaczew@fuw.edu.pl}
                   and J. Dudek$^{a,}$\footnote
                       {E-mail: jerzy@crnhp1.in2p3.fr}
        }

\vspace{3mm}
        {\it
              $^{a}$Centre de Recherches Nucl\'eaires,
                    IN$_2$P$_3$--CNRS/Universit\'e Louis Pasteur, Strasbourg I\\
                    F-67037 Strasbourg Cedex 2, France                        \\
              $^{b}$Institute of Theoretical Physics, Warsaw University       \\
                      ul. Ho\.za 69, PL-00681 Warsaw, Poland
        }
\end{center}

\vspace{5mm}
\hrule

\vspace{2mm}
\noindent{\bf Abstract}

We describe a method of solving the nuclear Skyrme-Hartree-Fock
problem by using a deformed Cartesian harmonic oscillator
basis. The complete list of expressions required to calculate
local densities, total energy, and self-consistent fields
is presented, and an implementation of the self-consistent symmetries
is discussed. Formulas to calculate matrix elements
in the Cartesian harmonic oscillator basis are derived
for the nuclear and Coulomb interactions.

\vspace{2mm}
\hrule

\vspace{2mm}


\section{Introduction}
\label{sec1}

Self-consistent methods have been used in the low-energy
nuclear structure studies over many years, and represent
a mature field with numerous successful applications, see
Refs.{\spc}\cite{RS80,Que78,Abe90a,Bak95,But96} for a review.  A
number of computer codes solving the nuclear Hartree-Fock (HF) problem
have already been developed. Two types of
effective nucleon-nucleon interactions have been mainly employed.  
Starting with the work of Vautherin and Brink \cite{Vau72}
many authors have applied the nuclear HF theory with the Skyrme effective 
interaction, while
the work of Gogny \cite{Gog73,Gog75b} initiated numerous studies
with the force which carries his name.

The methods employed to solve the HF equations depend mainly on
the effective force used and on the assumed symmetries of the
many-body wave functions.  For the solutions which allow at least
triaxial deformations, two different methods have been applied
for the two abovementioned effective interactions.  The first
one, used in conjuncture with the Skyrme interaction, is formulated in the
spatial coordinates and makes use of the finite-difference
\cite{Bon85}, or Fourier \cite{Bay86}, or spline-collocation
\cite{Uma91,Chi95} methods to approximate differential operators.  The
solution is then obtained by using the imaginary time evolution
operator \cite{Dav80}.

The second one, used for the finite-range Gogny interaction,
employs a truncated harmonic oscillator (HO) basis
\cite{Gog75,Gir83} and solves the problem either by an iterative
diagonalization of the mean-field Hamiltonian, or by the gradient-
\cite{Egi80} or the conjugate gradient- \cite{Egi95} methods.
A similar basis-expansion technique has also been recently used
\cite{Afa96} for
studies in the frame of the relativistic mean field theory,
for a review cf. Ref.{\spc}\cite{Rin96}.  The present study aims at
describing the method which incorporates the advantages of both
existing approaches, and combines the robustness of the Cartesian
HO basis with the simplicity of the Skyrme
interaction.

The methods using spatial coordinates have several
advantages.
First of all, various nuclear shapes can be easily
treated on the same footing; the same cubic lattice of
points in three spatial dimensions is suitable to
accommodate wave functions with, in principle, arbitrary deformations
restricted only by a specific
symmetrization of the lattice. This
allows easy studies of systems for which the deformation is not
{\it a priori} known, or is ill defined because of deformation
instabilities or a shape coexistence.  Secondly, using spatial coordinates
allows studies which address the asymptotic form of
nucleonic wave functions at large distances.  This is particularly
important for a precise description of weakly bound nuclei, where the use of 
spatial
coordinates is a necessity \cite{Dob84}.  Third, for the Skyrme zero-range
interaction, the mean fields are local (apart from a velocity
dependence) and can be easily programmed in the spatial
coordinates.  Last but not least, the treatment of wave
functions on large lattices (12$\times$12$\times$12 is a typical
example) is easily amenable to vectorization or parallelization
of the algorithm.

Methods using the HO basis have other
advantages.  Firstly, the basis provides a natural cut-off for many
operators which otherwise are unbound and require particularly
delicate treatment in the spatial coordinates.  This concerns
in particular the multipole moment and the angular momentum
operators which are often used as constraining operators.
For the corresponding constraints the solutions can become unstable when
the non-zero probability amplitudes (wave functions) move towards large 
distances as it is the case for e.g.{\spc}weakly bound nucleons. Secondly,
much smaller spaces are usually required to describe the nuclear
wave functions within a given precision.  Typically a basis of
about 300 HO wave functions is sufficient for most applications.
Third, the iterative diagonalization of the mean field
Hamiltonian can be used to find the self-consistent solutions, which
provides a rapidly converging algorithm, and, last but not
least, scalar or super scalar computers can also be used, because
the typical sizes of the information handled is smaller and the
performance is less dependent on the use of a vector processor.

The above listed advantages of a given method correspond often 
(although not always)
to the disadvantages of the other one, and both methods
described above are in this respect complementary.  Our
motivation to construct the Skyrme-Hartree-Fock code using the
Cartesian HO basis was based on the necessity to obtain a tool
which would allow rapid solutions for the nuclear superdeformed or
hyperdeformed rotating states for which the deformation is relatively
well known {\it a priori}.  Indeed, the method employed by us
gives a particularly fast, stable, robust, and simple algorithm
to solve such physical problems.

The paper is organized in the following way.
In Sec.{\spc}\ref{sec2} we present the HF method for the Skyrme
interaction, and in particular we discuss the local densities, total
energy, constraints, mean fields, and HF equations.
In Sec.{\spc}\ref{sec9} we describe the use of various symmetries
and in Sec.{\spc}\ref{sec3} the use of the Cartesian HO basis.
Sec.{\spc}\ref{sec4b} is devoted to the new method we use to
calculate the direct Coulomb potential. The method described
in the present study is implemented in the computer code \pr{HFODD},
published in the subsequent paper \cite{comcom2} which is referred to as II.

\section{Hartree-Fock method}
\label{sec2}

In this Section we present the complete set of formulas which
are used when solving the Skyrme-Hartree-Fock problem.

\subsection{Local densities}
\label{sec2c}

In the HF approximation, the total energy of a nuclear
system is, in general, a functional of the one-body non-local density
defined as
   \be\label{eq118a}
   \rho_\alpha(\bbox{r}\sigma,\bbox{r}'\sigma') =
   \langle\Psi|a^{\dagger}_{\cbox{r}'\sigma'\alpha}
                 a_{\cbox{r}\sigma\alpha}|\Psi\rangle.
   \ee
Here, $|\Psi\rangle$ is a many-body wave function, and 
$a^{\dagger}_{\mbox{\scriptsize\bf{{\it r}}}\sigma\alpha}$ and
$a_{\mbox{\scriptsize\bf{{\it r}}}\sigma\alpha}$ are
the operators creating and annihilating a neutron
($\alpha$=$n$) or a proton ($\alpha$=$p$) in the space-point
$\bbox{r}$ and having the projection of spin
$\sigma$=$\pm\frac{1}{2}$.

The non-local density (\ref{eq118a}) can be written \cite{EBG75} as
a sum of the scalar, $\rho_\alpha(\bbox{r},\bbox{r'})$, and
vector, $\bbox{s}_\alpha(\bbox{r},\bbox{r'})$ terms, defined by
   \be\label{eq118ba}
   \rho_\alpha(\bbox{r},\bbox{r}') 
  =\sum_{\sigma} \rho_\alpha(\bbox{r}\sigma,\bbox{r}'\sigma)
   \ee
and
   \be\label{eq118bb}
   \bbox{s}_\alpha(\bbox{r},\bbox{r}') 
  =\sum_{\sigma\sigma'} \rho_\alpha(\bbox{r}\sigma,\bbox{r}'\sigma')
   <\sigma'|\bbox{\sigma}|\sigma>
   \ee
as
   \be\label{eq118bc}
   \rho_\alpha(\bbox{r}\sigma,\bbox{r}'\sigma')
     = \frac{1}{2}\rho_\alpha(\bbox{r},\bbox{r}')\delta_{\sigma\sigma'}
     + \frac{1}{2}\bbox{s}_\alpha(\bbox{r},\bbox{r}')
                \cdot\bbox{\sigma}_{\sigma\sigma'}.
   \ee
Assuming that the total energy depends only on local
($\bbox{r}$=$\bbox{r}'$) densities,
and on their derivatives at $\bbox{r}$=$\bbox{r}'$ up to the second order,
one has to consider the following real nucleonic densities \cite{EBG75}:
\begin{itemize}

\item particle-density $\rho_\alpha(\bbox{r})$ and
      spin density $\bbox{s}_\alpha(\bbox{r})$,
   \bnll{eq205}
     \rho_\alpha(\bbox{r})
     &=& \rho_\alpha(\bbox{r},\bbox{r}),
                                               \label{eq205a} \\ \steplet
     \bbox{s}_\alpha(\bbox{r})
     &=& \bbox{s}_\alpha(\bbox{r},\bbox{r}),
                                                \label{eq205b}
   \enll

\item kinetic density $\tau_\alpha(\bbox{r})$ and
      vector kinetic density $\bbox{T}_\alpha(\bbox{r})$,
   \bnll{eq206}
     \tau_\alpha(\bbox{r})
     &=&
     \left[\bbox{\nabla}\cdot\bbox{\nabla}'
     \rho_\alpha(\bbox{r},\bbox{r}')\right]_{\bbox{r}=\bbox{r}'},
                                                 \label{eq206a} \\ \steplet
     \bbox{T}_\alpha(\bbox{r})
     &=&
     \left[\bbox{\nabla}\cdot\bbox{\nabla}'
     \bbox{s}_\alpha(\bbox{r},\bbox{r}')\right]_{\bbox{r}=\bbox{r}'},
                                                \label{eq206b}
   \enll

\item momentum density $\bbox{j}_\alpha(\bbox{r})$ and
      spin-current tensor $\tensor{J}_\alpha(\bbox{r})$,
   \bnll{eq207}
     \bbox{j}_\alpha(\bbox{r})&=&\frac{1}{2i}
     \left[(\bbox{\nabla}-\bbox{\nabla}')
     \rho_\alpha(\bbox{r},\bbox{r}')\right]_{\bbox{r}=\bbox{r}'},
                                                \label{eq207a} \\[2ex] \steplet
     J_{\mu\nu,\alpha}(\bbox{r})&=&\frac{1}{2i}
     \left[(\nabla_\mu-\nabla'_\mu)
     s_{\nu,\alpha}(\bbox{r},\bbox{r}')\right]_{\bbox{r}=\bbox{r}'}.
                                                \label{eq207b}
   \enll
\end{itemize}
In what follows we often omit the space argument $\bbox{r}$ of
local densities.

{}For each of the local densities we define the
isoscalar and isovector density. For example, the isoscalar ($\rho_0$)
and isovector ($\rho_1$) particle densities are defined as
   \be\label{eq110}
   \rho_0 = \rho_n + \rho_p \, , \quad
   \rho_1 = \rho_n - \rho_p.
   \ee
Following a traditional notation, for isoscalar densities
we sometimes omit the isospin index, for example,
$\rho$$\equiv$$\rho_0$.

\subsection{Hartree-Fock energy}
\label{sec2a}

In the Skyrme-HF approximation, the total energy ${\cal E}$
of a nucleus is given
as a sum of the kinetic, Skyrme, Coulomb, and pairing terms
   \be\label{eq504}
   {\cal E} = {\cal E}^{\text{kin}}
            + {\cal E}^{\text{Skyrme}}
            + {\cal E}^{\text{Coul}}
            + {\cal E}^{\text{pair}}.
   \ee
The kinetic energy of both protons and neutrons is given by the integral of 
the isoscalar kinetic density $\tau_0(\bbox{r})$,
cf.{\spc}Eqs.{\spc}(\ref{eq206a}) and (\ref{eq110}),
   \be\label{eq501}
   {\cal E}^{\text{kin}} = \frac{\hbar^2}{2m}\left(1-\frac{1}{A}\right)
                           \int d^3\bbox{r} \tau_0(\bbox{r}) ,
   \ee
where the standard factor $($1$-$$\frac{1}{A})$ provides a simple
approximation to the center-of-mass correction \cite{Bei75}
in terms of the number of nucleons $A$.

The Skyrme energy is the sum of the isoscalar ($t$=0) and isovector
($t$=1) terms, and is given as the integral of two energy
densities. The first of these densities,
${\cal H}^{\text{even}}_t(\bbox{r})$, depends on the time-even densities
$\rho_t$, $\tau_t$, and $\tensor{J}_t \equiv J_{\mu\nu,t}$, while the second 
one,
${\cal H}^{\text{odd}}_t (\bbox{r})$, depends on the time-odd  densities,
$\bbox{s}_t$, $\bbox{T}_t$, $\bbox{j}_t$
(see Ref.{\spc}\cite{J2D2b}), i.e.,
   \be\label{eq502}
   {\cal E}^{\text{Skyrme}} = \sum_{t=0,1} \int d^3\bbox{r}
                              \left({\cal H}^{\text{even}}_t(\bbox{r})
                            +       {\cal H}^{\text{odd}}_t (\bbox{r})\right),
   \ee
for
   \bnll{eq109}
   {\cal H}^{\text{even}}_t(\bbox{r})
  &\equiv& C_t^{\rho}            \rho_t^2
     +  C_t^{\Delta\rho}      \rho_t\Delta\rho_t
     +  C_t^{\tau}            \rho_t\tau_t
     +  C_t^{   J}            \tensor{J}_t^{\;2}
     +  C_t^{\nabla J}        \rho_t\bbox{\nabla}\cdot\bbox{J}_t,
                                                \label{eq109a}  \\ \steplet
   {\cal H}^{\text{odd}}_t(\bbox{r})
  &\equiv& C_t^{   s}            \bbox{s}_t^2
     +  C_t^{\Delta s}        \bbox{s}_t\cdot\Delta
                              \bbox{s}_t
     +  C_t^{   T}            \bbox{s}_t\cdot\bbox{T}_t
     +  C_t^{   j}            \bbox{j}_t^2
     +  C_t^{\nabla j}        \bbox{s}_t\cdot(\bbox{\nabla}\times\bbox{j}_t).
                                                \label{eq109b}
   \enll
In Eq.{\spc}(\ref{eq109a}), the square of the tensor density
is defined as $\tensor{J}_t^{\;2}$$\equiv$$\sum_{\mu\nu}J^{\;2}_{\mu\nu,t}$,
and its vector part $\bbox{J}_t$ is defined as
$J_{\lambda,t}$$\equiv$$\sum_{\mu\nu}\epsilon_{\lambda\mu\nu}J_{\mu\nu,t}$.

Expressions introduced via Eqs.\,(\ref{eq109}), relating the ten 
time-even coupling constants
$C_t^{\rho}$,
$C_t^{\Delta\rho}$,
$C_t^{\tau}$,
$C_t^{   J}$, and
$C_t^{\nabla J}$, and the ten time-odd
coupling constants
$C_t^{   s}$,
$C_t^{\Delta s}$,
$C_t^{   T}$,
$C_t^{   j}$, and
$C_t^{\nabla j}$,
to the standard parameters of the 
Skyrme interaction are given in
Ref.{\spc}\cite{J2D2b}.
Internally, the code \pr{HFODD} uses the coupling constants in the traditional
representation in which every term is a sum of the total density
squared, and of the sum of squares of neutron and proton densities.
Every such a term can also be written down in the isoscalar-isovector
representation used in Eqs.{\spc}(\ref{eq502}) and (\ref{eq109}).
For the simplest term depending on particle densities, the
correspondence is
    \be\label{eq580}
     C^\rho_{\text{tot}}(\rho_n+\rho_p)^2
  +  C^\rho_{\text{sum}}(\rho_n^2+\rho_p^2) =
     C^\rho_0(\rho_n+\rho_p)^2
  +  C^\rho_1(\rho_n-\rho_p)^2 ,
    \ee
which trivially leads to the following relations between both
sets of the coupling constants
   \bnll{eq582}
   C^\rho_{\text{tot}}  &=&    C^\rho_0  - C^\rho_1 ,
                                        \label{eq582a} \\ \steplet
   C^\rho_{\text{sum}}  &=&    2C^\rho_1            ,
                                        \label{eq582b}
   \enll
or equivalently
   \bnll{eq581}
   C^\rho_0 &=& {\textstyle{\frac{1}{2}}}
                 C^\rho_{\text{sum}} + C^\rho_{\text{tot}} ,
                                             \label{eq581a} \\ \steplet
   C^\rho_1 &=& {\textstyle{\frac{1}{2}}}
                 C^\rho_{\text{sum}}                      .
                                             \label{eq581b}
   \enll
Analogous relations hold for all other terms in the energy density.

The Coulomb energy is a sum of the direct and exchange contributions,
   \be\label{eq503}
   {\cal E}^{\text{Coul}} = {\cal E}^{\text{Coul}}_{\text{dir}}
                          + {\cal E}^{\text{Coul}}_{\text{exch}} ,
   \ee
for
   \bnll{eq505}
   {\cal E}^{\text{Coul}}_{\text{dir}}
       &=& \frac{e^2}{2} \int d\bbox{r}_1 d\bbox{r}_2
           \frac{\rho_p(\bbox{r}_1)\rho_p(\bbox{r}_2)}
                {|\bbox{r}_1-\bbox{r}_2|} , \label{eq505a} \\[2ex] \steplet
   {\cal E}^{\text{Coul}}_{\text{exch}}
       &=& \frac{e^2}{2} \int d\bbox{r}_1 d\bbox{r}_2
           \frac{\rho_p(\bbox{r}_2,\bbox{r}_1)\rho_p(\bbox{r}_1,\bbox{r}_2)}
                {|\bbox{r}_1-\bbox{r}_2|} , \label{eq505b}
   \enll\noindent%
where $\rho_p(\bbox{r})$ and $\rho_p(\bbox{r}_1,\bbox{r}_2)$
are the local and non-local proton densities, respectively, 
Eqs.{\spc}(\ref{eq205a}) and (\ref{eq118ba}),
and $e$ is the elementary charge.
By using expressions (\ref{eq505}) we approximate the charge density
by the proton density without taking into account
neither proton and neutron charge form factors \cite{Cam74} nor other
corrections \cite{Ber72} which turn out to be small.

In the algorithm,
the direct Coulomb energy is calculated as the trace of the proton density
matrix with the HO matrix elements of the Coulomb
potential (see Sec.{\spc}\ref{sec4b}).
The Coulomb energy $U^{\text{Coul}}(\bbox{r})$ is given by
   \be\label{eq513}
   U^{\text{Coul}}(\bbox{r})
        =  e^2 \int d\bbox{r}' \frac{\rho_p(\bbox{r}')}
                {|\bbox{r}-\bbox{r}'|} ,
   \ee
where, in order to express $U^{\text{Coul}}(\bbox{r})$ in units
of energy, the additional factor $e$ is added as compared to
the standard electrostatic expression.

The exchange Coulomb energy is calculated in
the Slater approximation \cite{Sla51,Tit74}:
   \be\label{eq506}
   {\cal E}^{\text{Coul}}_{\text{exch}}
        =  -\frac{3e^2}{4}\left(\frac{3}{\pi}\right)^{1/3} \int d\bbox{r}
                  \rho_p^{4/3}(\bbox{r}) .
   \ee

The term ${\cal E}^{\text{pair}}$ is equal to the average
value of the seniority pairing interaction \cite{RS80} calculated
in the BCS state:
   \be\label{eq510}
   {\cal E}^{\text{pair}} = - \sum_{\alpha=n,p}
           \frac{G_\alpha}{4} \left(\sum_{i}
                             u_{i,\alpha} v_{i,\alpha} \right)^2,
   \ee
where $G_n$ and $G_p$ are the neutron and proton pairing strengths,
respectively, taken from Ref.{\spc}\cite{DMS80}. Sums are performed 
separately over the neutron and proton
single-particle states, while the standard BCS occupation factors, 
$v_{i,\alpha}$ satisfy as always $v_{i,\alpha}^2+u_{i,\alpha}^2$=1.
Note that ${\cal E}^{\text{pair}}$ does not correspond to the
difference between the energies of unpaired and paired solutions,
because the occupation factors also influence other terms in the
total energy.
In the present version of the code \pr{HFODD}
the pairing correlations are included only within the
non-rotating case, see II.

\subsection{Constraints}
\label{sec2e}

The total energy (\ref{eq504}) can be minimized under
specific constraints. One of the most essential ones
is related to the so-called cranking approximation which
is equivalent to solving the time-dependent HF equations
in the laboratory frame or the HF equations in the rotating
frame \cite{RS80}. 
To distinguish between the energy operators and/or eigenvalues written either
in the laboratory frame or in the turning reference frame,
the latter ones are referred to as Routhians. This notion applies
also to situations when other constraints are taken into account.

In order to find a constrained minimum of energy
one has to find a minimum of the Routhian ${\cal E}'$ defined by
   \be\label{eq500}
   {\cal E}' = {\cal E}
             + {\cal E}^{\text{mult}}
             + {\cal E}^{\text{cran}}
             + {\cal E}^{\text{numb}},
   \ee
i.e., equal to the sum of the energy and the terms responsible for the
multipole, cranking, and particle-number constraints, respectively,
as defined below.

The multipole constraints are assumed in the standard
quadratic form \cite{FQK73}:
   \be\label{eq507}
   {\cal E}^{\text{mult}} = \sum_{\lambda\mu} C_{\lambda\mu}
                \left(\langle\hat Q_{\lambda\mu}\rangle
                           - \bar Q_{\lambda\mu}\right)^2 ,
   \ee
where $\langle\hat Q_{\lambda\mu}\rangle$ are the average values
of the mass-multipole-moment operators,
$\bar Q_{\lambda\mu}$ are the constraint values of the multipole
moments, and $C_{\lambda\mu}$ are the stiffness constants.

The cranking constraints aim at finding solutions corresponding
to non-zero angular momenta, and are assumed in a combined
form of the linear and quadratic constraints:
   \be\label{eq508}
   {\cal E}^{\text{cran}} = - \omega_J \langle\hat J_y\rangle
                            + C_J\left(\langle\hat J_y\rangle
                                         - \bar J_y\right)^2 ,
   \ee
where $\hat J_y$ is the operator of the component of the total
angular momentum along the $y$ axis, and $\bar J_y$ is the
corresponding target value.
The code \pr{HFODD} uses the $y$ axis as the cranking axis and
assumes that the $x$-$z$ plane as the conserved symmetry plane
(see below for more details).
For the combined constraint, the physical angular frequency $\omega$
is given by
   \be\label{eq512}
   \omega_y = \frac{\partial{\cal E}}
                 {\partial\langle\hat J_y\rangle}
               = \omega_J - 2C_J\left(\langle\hat J_y\rangle
                                         - \bar J_y\right) ,
   \ee
It turns out that a pure linear constraint
($C_J$=0)
leads to a much more stable convergence properties, and then
$\omega_y$=$\omega_J$.

The particle-number constraints ensure the correct average values
of neutron and proton numbers when the pairing option is used, and are assumed 
in the standard linear forms, i.e.,
   \be\label{eq511}
   {\cal E}^{\text{numb}} = - \lambda_n\langle\hat N_n\rangle
                            - \lambda_p\langle\hat N_p\rangle,
   \ee
where $\lambda_n$ and $\lambda_p$ are the neutron and proton Fermi
energies, and
$\hat N_n$ and $\hat N_p$ are the corresponding neutron and proton
particle-number operators. The particle number constraints
are taken into account only when the pairing correlations
are included (in the non-rotating case, at present). Otherwise
the numbers of particles are defined by selecting a given number of occupied
states explicitly (see II, Sec.{\spc}\ref{sec6e}).

\subsection{Hartree-Fock mean fields}
\label{sec2b}

Upon a variation of the total Routhian (\ref{eq500}) with respect to neutron
and proton single-particle wave functions, one obtains the
HF single-particle Routhian operators in the form:
    \bnll{eq515}
    \hspace{-2em}
    h_n' &=& -\frac{\hbar^2}{2m}\Delta
         +  \left(
            {\Gamma}^{\text{even}}_0 + {\Gamma}^{\text{odd}}_0
         +  {\Gamma}^{\text{even}}_1 + {\Gamma}^{\text{odd}}_1\right)
         + U^{\text{mult}} - \omega_y\hat J_y,
                                                            \\[2ex] \steplet
    \hspace{-2em}
    h_p' &=& -\frac{\hbar^2}{2m}\Delta
         +  \left(
            {\Gamma}^{\text{even}}_0 + {\Gamma}^{\text{odd}}_0
         -  {\Gamma}^{\text{even}}_1 - {\Gamma}^{\text{odd}}_1\right)
         + U^{\text{Coul}} + U^{\text{mult}} - \omega_y\hat J_y.
    \enll
The momentum dependent potentials $\Gamma$ are given
by \cite{EBG75,J2D2b}
   \bnll{eq209}
   {\Gamma}^{\text{even}}_t
    &=& -\bbox{\nabla}\cdot \left[M_t(\bbox{r})\bbox{\nabla}\right]
     +  U_t(\bbox{r})
     +  \frac{1}{2i}\Big(\tensor{\nabla\sigma}\cdot\tensor{B}_t(\bbox{r})
                        + \tensor{B}_t(\bbox{r})\cdot\tensor{\nabla\sigma}
                    \Big),
                                                          \\[2ex] \steplet
   {\Gamma}^{\text{odd}}_t
    &=& -\bbox{\nabla}\cdot
         \left[\Big(\bbox{\sigma}\cdot\bbox{C}_t(\bbox{r})\Big)
         \bbox{\nabla}\right]
     +  \bbox{\sigma}\cdot\bbox{\Sigma}_t(\bbox{r})
     +  \frac{1}{2i}\Big(\bbox{\nabla}\cdot\bbox{I}_t(\bbox{r})
                        + \bbox{I}_t(\bbox{r})\cdot\bbox{\nabla}\Big).
   \enll
In Eqs.{\spc}(\ref{eq515}), $U^{\text{mult}}$
represents the terms originating
from the multipole constraints:
   \be\label{eq514}
   U^{\text{mult}} =
       2\sum_{\lambda\mu} C_{\lambda\mu}
                \left(\langle\hat Q_{\lambda\mu}\rangle
                           - \bar Q_{\lambda\mu}\right)
                           \hat Q_{\lambda\mu}.
   \ee
With pairing correlations taken into account, the single-particle Routhians
should in principle also contain the terms $-$$\lambda_\alpha\hat N_\alpha$
originating from the particle number constraint (\ref{eq511}).
However, usually these terms are kept aside and included in the
equations explicitly. Such a convention allows keeping
for the single-particle Routhians the convenient
standard energy scale.

The functions defining the HF potentials (\ref{eq209})
depend on the local densities (\ref{eq205})--(\ref{eq207}),
and read \cite{J2D2b}
   \bnll{eq210}
   U_t
       &=& 2C_t^{\rho}           \rho_t
        +  2C_t^{\Delta\rho}      \Delta\rho_t
        +  C_t^{\tau}            \tau_t
        +  C_t^{\nabla J}        \bbox{\nabla}\cdot\bbox{J}_t
        +  U_t',
                                             \label{eq210a} \\   \steplet
   \bbox{\Sigma}_t
       &=& 2C_t^{   s}           \bbox{s}_t
        +  2C_t^{\Delta s}        \Delta\bbox{s}_t
        +  C_t^{   T}            \bbox{T}_t
        +  C_t^{\nabla j}        \bbox{\nabla}\times\bbox{j}_t,
                                             \label{eq210b} \\   \steplet
   M_t
       &=& C_t^{\tau}            \rho_t,     \label{eq210c} \\   \steplet
   \bbox{C}_t
       &=& C_t^{   T}            \bbox{s}_t, \label{eq210d} \\   \steplet
   \tensor{B}_t
       &=& 2C_t^{   J}           \tensor{J}_t
        -  C_t^{\nabla J}        \tensor{\nabla}\rho_t,
                                             \label{eq210e} \\   \steplet
   \bbox{I}_t
       &=& 2C_t^{   j}           \bbox{j}_t
        +  C_t^{\nabla j}        \bbox{\nabla}\times\bbox{s}_t.
                                             \label{eq210f}
   \enll
The tensor gradient operators in Eqs.{\spc}(\ref{eq209}) and
(\ref{eq210e}) are defined \cite{EBG75} as
$(\tensor{\nabla\sigma})_{\mu\nu}$=$\nabla_\mu\sigma_\nu$ and
$\nabla_{\mu\nu}$$=$$\sum_{\lambda}\epsilon_{\mu\nu\lambda}\nabla_\lambda$.

In Eq.{\spc}(\ref{eq210a}) the term $U_t'$ represents the rearrangement
terms resulting from the density dependence of the coupling constants.
In standard parametrizations of the Skyrme forces, which are implemented
in the code \pr{HFODD}, only the $C_t^{\rho}$ and $C_t^s$ coupling
constants depend on the isoscalar density $\rho_0$, and then
   \be\label{eq211}
    U_t' = \delta_{t0}\sum_{t'=0,1}\left(
           \frac{\partial C_{t'}^{\rho}}{\partial\rho_0}\rho_{t'}^2
         + \frac{\partial C_{t'}^{   s}}{\partial\rho_0}\bbox{s}_{t'}^2\right).
   \ee

\subsection{Hartree-Fock equations}
\label{sec2d}

The eigenequations for the HF single-particle Routhians (\ref{eq515}) are
called HF equations, and read
   \be\label{eq516}
    h_\alpha'\psi_{i,\alpha}(\bbox{r}\sigma)
       = e_{i,\alpha}'\psi_{i,\alpha}(\bbox{r}\sigma) .
   \ee
The single-particle index $i$ comprises all nucleonic quantum numbers. 
We use the convention that this index has
different values for both members within the
time-reversed, or signature-reversed, or simplex-reversed couples
(see Sec.{\spc}\ref{sec9}), i.e., every single-particle state is
considered separately, and the sums over $i$ are performed over
{\em all} neutron or proton single-particle states.  This is in
contrast to some formulations where the sums are performed over
only one-half of states which have one common value of, e.g., the
signature, or one sign of the spin projection.  [Note, for
example, the factor $\frac{1}{4}$ in Eq.{\spc}(\ref{eq510}).]

The standard way of assessing the quality of convergence of the
HF equations is to compare the value of energy ${\cal E}$
calculated from Eqs.{\spc}(\ref{eq504})--(\ref{eq510}) with that
calculated by using the sum of the single-particle energies
   \be\label{eq595}
   {\cal E}^{\text{s.p.}} = \sum_{i,\alpha} v_{i,\alpha}^2 e_{i,\alpha}.
   \ee
A connection between these two energies exists because
${\cal E}^{\text{s.p.}}$ is related to the average value of the
single-particle Routhian, which for a density independent
interaction and with no constraints is trivially equal to the
average kinetic energy plus twice the average two-body
interaction energy \cite{RS80}.
In the realistic case of the density-dependent 
interactions, this simple relation
has to be modified and one defines the equivalent energy
$\tilde{\cal E}$ by
   \be\label{eq596}
   \tilde{\cal E} = \thalf{\cal E}^{\text{s.p.}}
                  + \thalf{\cal E}^{\text{kin}}
                  +       {\cal E}^{\text{pair}}
                  -       {\cal E}^{\text{rear}}
                  + \third{\cal E}^{\text{Coul}}_{\text{exch}}
                  -       {\cal E}^{\text{mult}}_{\text{corr}}
                  -       {\cal E}^{\text{cran}}_{\text{corr}}
   \ee
Here, the rearrangement energy ${\cal E}^{\text{rear}}$ results
from the density dependence of the Skyrme interaction and is
equal to one half of the average value of rearrangement potential
Eq.{\spc}(\ref{eq211}),
   \be\label{eq597}
   {\cal E}^{\text{rear}} = \thalf\int d^3\bbox{r} U'_0\rho_0.
   \ee
Similarly, the Coulomb exchange energy (\ref{eq506}) can be
considered as resulting from a zero-order interaction term
depending on the density as $\rho_p^{-2/3}$.  Hence the resulting
rearrangement term in Eq.{\spc}(\ref{eq596}) is equal to the
Coulomb exchange energy with the factor $\third$.  Finally, the
corrections resulting from the constraints read
   \be\label{eq600}
   {\cal E}^{\text{mult}}_{\text{corr}} = \sum_{\lambda\mu} C_{\lambda\mu}
                \left(\langle\hat Q_{\lambda\mu}\rangle
                           - \bar Q_{\lambda\mu}\right)
                           \langle\hat Q_{\lambda\mu}\rangle
   \ee
for the multipole constraints (\ref{eq514}), and
   \be\label{eq598}
   {\cal E}^{\text{cran}}_{\text{corr}} = -\thalf\omega_y\langle\hat J_y\rangle
   \ee
for the cranking constraint with $\omega_y$ given by
Eq.{\spc}(\ref{eq512}).  Since the term corresponding to the
particle number constraint is not included in the Routhian, the
correction for this constraint does not appear in $\tilde{\cal E}$.

The difference between the energies $\tilde{\cal E}$ and ${\cal E}$
is exactly equal to zero when the densities and fields do not change
from one iteration to the next one. Hence their difference
   \be\label{eq599}
   \delta{\cal E} = \tilde{\cal E} - {\cal E}
   \ee
provides a useful measure of the quality of convergence, and is called
the stability of the HF energy. In practice, ${\cal E}$ approaches
faster the final value of the total energy energy than does
$\tilde{\cal E}$.

\subsection{Hartree-Fock densities}
\label{sec2f}

In terms of the Routhian eigenfunctions, the non-local density matrix
is given by
   \be\label{eq517}
   \rho_\alpha(\bbox{r}\sigma,\bbox{r}'\sigma') =
      \sum_{i} v_{i,\alpha}^2
                 \psi_{i,\alpha}(\bbox{r}\sigma)
                 \psi_{i,\alpha}^*(\bbox{r}'\sigma') .
   \ee
The occupation factors $v_{i,\alpha}^2$ are determined
through the standard BCS
procedure (when the pairing correlations are taken into account
for the non-rotating case of the code \pr{HFODD}),
or are equal to 1 or 0 according to the chosen configuration
(when the pairing correlations are not used).

The energy densities (\ref{eq109}) depend on the six real densities
$\rho_t$, $\tau_t$, $\bbox{s}_t$,
$\bbox{T}_t$, $\bbox{j}_t$, and $\tensor{J}_t$. Each of them has
the isoscalar ($t$=0) and isovector ($t$=1) form, or equivalently,
the neutron and proton form. Moreover, in Eqs.{\spc}(\ref{eq109})
and (\ref{eq210}) there also appear six secondary densities which are
some particular derivatives of the six basic ones, i.e.,
$\Delta\rho_t$, $\bbox{\nabla}\cdot\bbox{J}_t$, $\Delta\bbox{s}_t$,
$\bbox{\nabla}\rho_t$, $\bbox{\nabla}\times\bbox{j}_t$, and
$\bbox{\nabla}\times\bbox{s}_t$. In principle, they can be
calculated by an explicit numerical derivation of the six basic
densities. However, the organization of the code \pr{HFODD} allows
calculating them directly at a negligible expense of the CPU time.
This is so, because the principal numerical effort must anyhow be devoted
to summing up the contributions of the HO basis states to the
HF eigenstates at every point in space
(see Sec.{\spc}\ref{sec3c}). Once this is accomplished,
the remaining operations are very rapid.

The scheme used in the code \pr{HFODD} is therefore the following.
{}First, at every point of space, $\bbox{r}$, and for every spin value $\sigma$,
the values of the HF wave functions,
their derivatives
in three directions, and their Laplasians are calculated. At this point it is 
convenient to combine the value of each wave function and its
three derivatives into one four-vector. To this effect we introduce
the symbolic notation for derivatives, $\nabla_{\hat\mu}$, which are
distinguished
by hatted indices taking values ${\hat\mu}$=0, 1, 2, 4. Then,
${\hat\mu}$=0 corresponds to the unity operator (no derivative involved), 
and the three usual indices
correspond to the usual gradient in space, i.e.,
$\nabla_{\hat\mu}$$\equiv$$(1,\bbox{\nabla})$. By using this
notation we define the two following generic matrices of densities:
   \bnll{eq518}
      D^{qq'}_{{\hat\mu}{\hat\nu},\alpha} &=& \sum_{i} v_{i,\alpha}^2
 \left[\nabla_{\hat\mu}\psi_{i,\alpha}
                                   (\bbox{r}\sigma)\right]
 \left[\nabla_{\hat\nu}\psi_{i,\alpha}^*
                                   (\bbox{r}\sigma')\right] ,
                                         \label{eq518a} \\ \steplet
      L^{qq'}_{\alpha}                    &=& \sum_{i} v_{i,\alpha}^2
                       \psi_{i,\alpha}  (\bbox{r}\sigma)
                 \Delta\psi_{i,\alpha}^*(\bbox{r}\sigma') ,
                                         \label{eq518b}
   \enll
where indices $q$ and $q'$ denote the signs of spin variables
$\sigma$ and $\sigma'$, respectively,
i.e., $q$=$+$ for $\sigma$=$+\frac{1}{2}$ and
$q$=$-$ for $\sigma$=$-\frac{1}{2}$.

The local densities required for the
calculation of energies and fields have to be determined separately
for neutrons and protons.
Below we omit for simplicity the isospin indices
and denote by $\Re$ and $\Im$ the real and imaginary parts.
Then the required 34 real densities are given in terms of
64 complex or real densities
$D^{qq'}_{{\hat\mu}{\hat\nu}}$ and 4 complex densities
$L^{qq'}$ by the following expressions:
\bi
\item scalar densities
   \bn
   \hspace{-3em} \rho &=&
                            D^{++}_{ 0  0 } + D^{--}_{ 0  0 } ,
                                    \label{eq519a} \\
   \hspace{-3em} \tau &=&
             \sum_\mu \left(D^{++}_{\mu\mu} + D^{--}_{\mu\mu}\right) ,
                                    \label{eq519b} \\
   \hspace{-3em} \Delta\rho &=&
      \phantom{-}2\Re \left(L^{++}          + L^{--}         \right)
               + 2\tau ,
                                    \label{eq519c} \\
   \hspace{-3em} \bbox{\nabla}\cdot\bbox{J} &=&
               - 2\Im \left(D^{+-}_{ 2  3 } - D^{+-}_{ 3  2 }\right)
               - 2\Re \left(D^{+-}_{ 3  1 } - D^{+-}_{ 1  3 }\right)
               + 2\Im \left(D^{++}_{ 2  1 } - D^{--}_{ 2  1 }\right) ,
                                    \label{eq519d}
   \en

\item vector densities
   \bnll{eq520}
    s_1 &=&
      \phantom{-}2\Re  D^{+-}_{ 0  0 } ,
                                    \label{eq520a} \\ \steplet
    s_2 &=&
               - 2\Im  D^{+-}_{ 0  0 } ,
                                    \label{eq520b} \\ \steplet
    s_3 &=&
                       D^{++}_{ 0  0 } - D^{--}_{ 0  0 } ,
                                    \label{eq520c}
   \enll
   \bnll{eq521}
    T_1 &=&
      \phantom{-}2\Re \sum_\mu       D^{+-}_{\mu\mu} ,
                                    \label{eq521a} \\ \steplet
    T_2 &=&
               - 2\Im \sum_\mu       D^{+-}_{\mu\mu} ,
                                    \label{eq521b} \\ \steplet
    T_3 &=&
                      \sum_\mu \left(D^{++}_{\mu\mu} - D^{--}_{\mu\mu}\right) ,
                                    \label{eq521c}
   \enll
   \bnll{eq522}
    \Delta s_1 &=&
                 2\Re          \left(L^{+-}         + L^{-+}         \right)
               + 2T_1 ,
                                    \label{eq522a} \\ \steplet
    \Delta s_2 &=&
                 2\Im          \left(L^{+-}         - L^{-+}         \right)
               + 2T_2 ,
                                    \label{eq522b} \\ \steplet
    \Delta s_3 &=&
      \phantom{\Re}2             \left(L^{++}         - L^{--}         \right)
               + 2T_3 ,
                                    \label{eq522c}
   \enll
   \bn
    \nabla\rho_\mu &=&
                 2\Re          \left(D^{++}_{\mu 0 } + D^{--}_{\mu 0 }\right) ,
                                    \label{eq523a} \\
             j_\mu &=&
       \phantom{2}\Im          \left(D^{++}_{\mu 0 } + D^{--}_{\mu 0 }\right) ,
                                    \label{eq523b}
   \en
   \bnll{eq524}
    (\bbox{\nabla}\times\bbox{s})_1 &=&
      \phantom{-}2\Re          \left(D^{++}_{ 2  0 } - D^{--}_{ 2  0 }\right)
               + 2\Im          \left(D^{+-}_{ 3  0 } + D^{+-}_{ 0  3 }\right) ,
                                    \label{eq524a} \\ \steplet
    (\bbox{\nabla}\times\bbox{s})_2 &=&
      \phantom{-}2\Re          \left(D^{+-}_{ 3  0 } + D^{+-}_{ 0  3 }\right)
               - 2\Re          \left(D^{++}_{ 1  0 } - D^{--}_{ 1  0 }\right) ,
                                    \label{eq524b} \\ \steplet
    (\bbox{\nabla}\times\bbox{s})_3 &=&
               - 2\Im          \left(D^{+-}_{ 1  0 } + D^{+-}_{ 0  1 }\right)
               - 2\Re          \left(D^{+-}_{ 2  0 } + D^{+-}_{ 0  2 }\right) ,
                                    \label{eq524c}
   \enll
   \bnll{eq525}
    (\bbox{\nabla}\times\bbox{j})_1 &=&
      \phantom{-}2\Im          \left(D^{++}_{ 3  2 } + D^{--}_{ 3  2 }\right) ,
                                    \label{eq525a} \\ \steplet
    (\bbox{\nabla}\times\bbox{j})_2 &=&
               - 2\Im          \left(D^{++}_{ 3  1 } + D^{--}_{ 3  1 }\right) ,
                                    \label{eq525b} \\ \steplet
    (\bbox{\nabla}\times\bbox{j})_3 &=&
      \phantom{-}2\Im          \left(D^{++}_{ 2  1 } + D^{--}_{ 2  1 }\right) ,
                                    \label{eq525c}
   \enll

\item tensor density
   \bnll{eq526}
    J_{\mu1} &=&
                  \Im          \left(D^{+-}_{\mu 0 } - D^{+-}_{ 0 \mu}\right) ,
                                    \label{eq526a} \\ \steplet
    J_{\mu2} &=&
                  \Re          \left(D^{+-}_{\mu 0 } - D^{+-}_{ 0 \mu}\right) ,
                                    \label{eq526b} \\ \steplet
    J_{\mu3} &=&
                  \Im          \left(D^{++}_{\mu 0 } - D^{--}_{\mu 0 }\right) .
                                    \label{eq526c}
   \enll
\ei


\section{Self-consistent symmetries}
\label{sec9}

\subsection{Simplex}
\label{sec9a}

The reader is referred to
Ref.{\spc}\cite{DDRW} for a detailed discussion of various
possible choices of conserved symmetries and of the implications
thereof. In the present version of the code \pr{HFODD} we assume
$x$-$z$ plane as {\em a priori} the only symmetry plane of the HF states.
Such an assumption has been dictated exclusively by the
fact that the CPU time for a typical run is markedly lower when symmetries
are assumed. However, in physical applications one usually requires that
all symmetries are determined by the physical system itself as a result
of the self-consistency condition and not by a pre-supposition. We find that
the assumption of a single symmetry plane provides an acceptable compromise
between the above conflicting criteria.

As a consequence of the above assumption, the $y$-simplex operator, defined by
   \be\label{eq527}
   \hat{S}_y = \hat P\exp\left(-i\pi\hat J_y\right) ,
   \ee
where $\hat P$ is the parity operator,
commutes with the single-particle Routhian operators
(\ref{eq515}), i.e.,
   \be\label{eq529}
   \left[h_n',\hat{S}_y\right] = \left[h_p',\hat{S}_y\right] = 0 ,
   \ee
and that the many-body wave function $|\Psi\rangle$ is an eigenstate
of $\hat{S}_y$:
   \be\label{eq591}
   \hat{S}_y|\Psi\rangle = S|\Psi\rangle.
   \ee
There are some technical advantages of choosing of the $y$-simplex as the 
conserved symmetry. One of them is related to the convenient phase relations
which are discussed below; another one consists in the 
fact that the average values of multipole moments:
   \be\label{eq590}
   Q_{\lambda\mu} = \langle\hat Q_{\lambda\mu}\rangle
               \equiv \langle\Psi|\hat Q_{\lambda\mu}|\Psi\rangle
   \ee
are real, i.e.,
   \be\label{eq528}
        Q_{\lambda\mu}^* = Q_{\lambda\mu}.
   \ee
This property results from combining the standard time-invariance
condition \cite{Var89},
   \be\label{eq539}
        \hat Q_{\lambda\mu}^*
     =  (-1)^{\mu}\hat Q_{\lambda,-\mu} ,
   \ee
with the transformation law under the operation of the $y$-simplex:
   \be\label{eq540}
        \hat{S}_y^{\dagger}\hat Q_{\lambda\mu}\hat{S}_y
     =  (-1)^{\mu}\hat Q_{\lambda,-\mu}.
   \ee
Equations (\ref{eq528})--(\ref{eq539}) allow to consider only (real) multipole 
moments with
non-negative magnetic components, $\mu\geq0$. Apart from a restriction
to real values, all multipole moments can be non-zero.
In particular, deformations corresponding to
four magnetic components, $\mu$=0, 1, 2, and 3,
of the octupole moment can be simultaneously present.

The choice of the $y$-simplex as a conserved symmetry
requires that the nuclear rotation (cranking) takes place around the $y$ axis. 
In the following
we refer to the $y$-simplex as the simplex, and keep in mind
that this axis corresponds to the rotation axis.

Because
   \be\label{eq605}
   \hat{S}_y^2 =  (-1)^A,
   \ee
the simplex has four eigenvalues:
$S$=$\pm{1}$ and $S$=$\pm{i}$ for even and odd numbers of particles,
respectively.
Due to the commutation relations (\ref{eq529}), the single-particle Routhians
do not have non-zero matrix elements between states belonging
to different eigenvalues of $\hat{S}_y$.
The equations for the single-particle eigenvalues can, therefore, be solved 
separately for both values of the single-particle simplex
$s$=$\pm{i}$, and the single-particle wave functions
$\psi_{i,\alpha}(\bbox{r}\sigma)$ are labeled
by these values of the simplex, i.e.,
   \be\label{eq538}
   \hat{S}_y \psi_{i,\alpha}(\bbox{r}\sigma)
       = s_i \psi_{i,\alpha}(\bbox{r}\sigma) .
   \ee
In the above relation we do not introduce $s_i$ as an extra label of the 
wave function since, according to our convention, all quantum numbers 
characterizing the single-particle states are contained within the index $i$. 
Throughout the code, input data, and output files we associate
the sign ``+'' with the eigenvalue $s$=$+i$ of the simplex
operator $\hat{S}_y$, and the
sign ``$-$'' with the eigenvalue $s$=$-i$.

\subsection{Time-reversal}
\label{sec9b}

The code \pr{HFODD} solves the HF equations with or without
the time-reversal symmetry imposed, depending on the value of the input 
parameter {\tt{}IROTAT} (see II, Sec.{\spc}\ref{sec6f}).
However, the antiunitary time-reversal operator $\hat{T}$
   \be\label{eq530}
   \hat{T} = -i\sigma_y\hat{K}_0 \quad,\quad \hat{T}^2 = (-1)^A,
   \ee
where $\hat{K}_0$ is the complex conjugation in coordinate space,
is used throughout to establish
the phases of the HO-basis simplex eigenstates, Sec.{\spc}\ref{sec3b}.

Since the time-reversal operator
commutes with the simplex operator,
   \be\label{eq533}
   \hat{T}\hat{S}_y = \hat{S}_y\hat{T}
   \ee
the time-reversed single-particle
states belong to opposite simplex eigenvalues.
(Recall that the antilinear time-reversal operator changes
the sign of the imaginary simplex eigenvalue.)
Therefore, only matrix elements for one of the eigenvalues,
say $s$=+$i$, have to be evaluated, while those for $s$=$-i$
can be obtained by the time-reversal. One has only to keep
track of the time-even and time-odd terms in the Routhians.
When both kinds of terms are non-zero, as is for the case of
rotation, the Routhians for both $s$ eigenvalues have to be
diagonalized. However, when the time-reversal symmetry is present (no rotation),
only one diagonalization is sufficient.

\subsection{Parity and signature}
\label{sec9c}

The code \pr{HFODD} solves the HF equations with or without
the parity symmetry imposed, depending on the value of the parameter
{\tt{}ISIGNY} chosen in the input data
(see II, Sec.{\spc}\ref{sec6f}).
Since the simplex $\hat{S}_y$, Eq.{\spc}(\ref{eq527}),
is in the present implementation
always a conserved symmetry, the case of conserved parity
is in fact realized by requiring that the $y$-signature
(signature for short), defined by
   \be\label{eq534}
   \hat{R}_y = \exp\left(-i\pi\hat J_y\right) ,
   \ee
is a conserved symmetry.
Then, the HF single-particle states are in addition labeled by the
signature eigenvalues, $r$=$\pm{i}$,
   \be\label{eq535}
   \hat{R}_y \psi_{i,\alpha}(\bbox{r}\sigma)
       = r_i \psi_{i,\alpha}(\bbox{r}\sigma) ,
   \ee
while the simplex is a product of parity $\pi$ and signature $r$,
$s$=$\pi{r}$, or equivalently the parity is a product of simplex
and signature, $\pi$=$sr^*$. As before, the signature quantum number $r_i$
is not attached explicitly as a label of the wave function.
The $s$=$+i$ block splits into two parity-signature
blocks ($\pi$,$r$)=(+1,$+i$) and ($-$1,$-i$), and similarly
the $s$=$-i$ block splits into
($-$1,$+i$) and (+1,$-i$).
These four blocks can be diagonalized separately.
Throughout the code, input data, and output files we associate
the sign ``+'' with the eigenvalue $r$=$+i$
of the signature operator $\hat{R}_y$, and the
sign ``$-$'' with the eigenvalue $r$=$-i$.

In the case of
conserved parity and signature, the multipole moments
$Q_{\lambda\mu}$ vanish for odd values of $\lambda$,
which is a consequence of the transformation law for
multipole moments:
   \be\label{eq541}
        \hat{P}^{\dagger}\hat Q_{\lambda\mu}\hat{P}
     =  (-1)^{\lambda}\hat Q_{\lambda\mu}.
   \ee

\subsection{T-simplexes}
\label{sec9d}

In addition to the two important cases of either the simplex $\hat{S}_y$ 
alone appearing as a conserved quantity,
or of the parity and signature $\hat{R}_y$  being conserved,
one may have two other cases of conserved
symmetries. These symmetries cannot give any new quantum
numbers simultaneously with the simplex $s$,
because none of the other simplexes,
$\hat{S}_x$ and $\hat{S}_z$, or signatures,
$\hat{R}_x$ or $\hat{R}_z$, commutes with $\hat{S}_y$.
However, we may still have conserved antilinear operators,
which do not provide quantum numbers, but restrict
spatial properties of solutions. The code \pr{HFODD}
can treat two such a symmetries \cite{DDRW}, the $x$-simplex$^T$
and the $z$-simplex$^T$,
   \bnll{eq536}
   \hat{S}_x^T &=& \hat{T}\hat{P}\exp\left(-i\pi\hat{J}_x\right)
                         \;,\quad \left(\hat{S}_x^T\right)^2 = 1  ,
                                               \label{eq536a} \\ \steplet
   \hat{S}_z^T &=& \hat{T}\hat{P}\exp\left(-i\pi\hat{J}_z\right)
                         \;,\quad \left(\hat{S}_z^T\right)^2 = 1  .
                                               \label{eq536b}
   \enll
These symmetries are requested by activating in the input data
(see II, Sec.{\spc}\ref{sec6f}) the
options {\tt{}ISIMTX} and {\tt{}ISIMTZ}, respectively.
The product of these two symmetries equals to the $y$-signature,
   \be\label{eq537}
   \hat{S}_x^T \hat{S}_z^T = \hat{R}_y .
   \ee
Therefore, if the Hamiltonian commutes with both of them, the
signature (and hence parity) must also 
be conserved.

The two $T$-simplexes transform the multipole moments in the
following way:
   \bnll{eq542}
        (\hat{S}_x^T)^{\dagger}\hat Q_{\lambda\mu}\hat{S}_x^T
     &=&  (-1)^{\mu}\hat Q_{\lambda\mu} ,
                                               \label{eq542a} \\ \steplet
        (\hat{S}_z^T)^{\dagger}\hat Q_{\lambda\mu}\hat{S}_z^T
     &=&  (-1)^{\lambda}\hat Q_{\lambda,-\mu} ,
                                               \label{eq542b}
   \enll
which combined with the time-reversal and $y$-simplex transformation
laws (\ref{eq539}) and (\ref{eq540}) gives, respectively, the
following properties
of multipole moments:
   \bnll{eq543}
        Q_{\lambda\mu} &=&  (-1)^{\mu}        Q_{\lambda\mu},
                                               \label{eq543a} \\ \steplet
        Q_{\lambda\mu} &=&  (-1)^{\lambda+\mu}Q_{\lambda\mu}.
                                               \label{eq543b}
   \enll

Therefore, commutation of $\hat{S}_x^T$ with the Hamiltonian 
enforces the vanishing of all
odd magnetic components of all multipole moments.
Nevertheless, with broken
parity, the even magnetic
components of odd multipoles, like $Q_{30}$ and $Q_{32}$, can be nonzero.
On the other hand, conservation of $\hat{S}_z^T$ enforces
the vanishing of odd magnetic components of even multipoles,
and the vanishing of even magnetic components of odd multipoles.
Therefore, with broken parity, the moments $Q_{31}$ and $Q_{33}$
can be nonzero. Of course, conservation of both $T$-simplexes
leaves nonzero only even magnetic components of even multipoles.
Table \ref{tab1} summarizes the properties of multipole moments under
different symmetry conditions. One should note that the broken
parity may lead to the system which has the center of mass shifted
from the origin,
$Q_{10}$$\neq$0 and/or $Q_{11}$$\neq$0,
while for the both $T$-simplexes broken it may in addition
lead to a shape which is not in the principal axes of the quadrupole
tensor, $Q_{21}$$\neq$0. In order to avoid physically meaningless
``deformations'' which in fact correspond to a shift or to a rotation
of the whole system, one has to use suitable constraints,
see Sec.{\spc}\ref{sec2e} and II -- Sec.{\spc}\ref{sec6r},
to enforce conditions
$Q_{10}$=$Q_{11}$=$Q_{21}$=0 in cases where the non-zero
values are, in principle, allowed by broken symmetries.

Condition (\ref{eq537}) means that there exist a possibility
\cite{DDRW} of breaking both antilinear symmetries, $\hat{S}_x^T$
and $\hat{S}_z^T$, and still conserving the signature (and hence
parity).  Such a possibility is, however, excluded from the
present version of the code.  This is so because, it would have
corresponded to a state with vanishing octupole moments, and
nevertheless non-zero moments $Q_{41}$ and $Q_{43}$ present in
the principal-axes reference frame ($Q_{21}$=0).  The shapes of
that type are very exotic indeed,
cf.{\spc}Ref.{\spc}\cite{Roh96}, and probably should be studied
after an investigation of those corresponding to non-zero moments
$Q_{41}$ and $Q_{43}$ induced by the non-zero octupole
deformation.  The latter possibility is allowed in the code
\pr{HFODD}.


\section{Cartesian harmonic oscillator basis}
\label{sec3}

The code \pr{HFODD} solves the HF equations by expanding
the single-particle wave functions $\psi_i(\bbox{r}\sigma)$
onto the deformed HO wave functions $\psi_{n_xn_yn_z,s_z}(\bbox{r}\sigma)$ 
in the Cartesian coordinates, i.e.,
   \be\label{eq544}
   \psi_i(\bbox{r}\sigma)
           = \sum_{n_x=0}^{N_x}
             \sum_{n_y=0}^{N_y}
             \sum_{n_z=0}^{N_z}
             \sum_{s_z=-\frac{1}{2},\frac{1}{2}}
             A_i^{n_xn_yn_z,s_z}\psi_{n_xn_yn_z,s_z}(\bbox{r}\sigma).
   \ee
Here $N_x$, $N_y$, and $N_z$ are the maximum numbers of the HO
quanta corresponding to the three Cartesian directions. However, as discussed
in Sec.{\spc}\ref{sec5a} of II,
the sums over $n_x$, $n_y$, and $n_z$ are performed over
the grid of points which form a pyramid rather than a cube.

The HO wave functions have the standard form
   \be\label{eq545}
   \psi_{n_xn_yn_z,s_z}(\bbox{r}\sigma)
     = \psi_{n_x}(x)\psi_{n_y}(y)\psi_{n_z}(z)\delta_{s_z\sigma},
   \ee
where
   \be\label{eq546}
   \psi_{n_\mu}(x_\mu) = b_\mu^{\frac{1}{2}}H_{n_\mu}^{(0)}(\xi_\mu)
                         e^{-\frac{1}{2}\xi_\mu^2},
   \ee
and $\xi_\mu$=$b_\mu{}x_\mu$ are dimensionless variables scaled
by the oscillator constants 
   \be\label{eq549}
       b_\mu=\sqrt{m\omega_\mu/\hbar}.
   \ee
Polynomials $H_{n}^{(0)}(\xi)$ are proportional to the standard
Hermite orthogonal polynomials $H_{n}(\xi)$ \cite{Abr70},
   \be\label{eq557}
    H_{n}^{(0)}(\xi) = \left(\sqrt{\pi}2^nn!\right)^{-\frac{1}{2}}
                       H_{n}(\xi),
   \ee
 and normalized as
   \be\label{eq547}
   \int_{-\infty}^{\infty} d\xi H_n^{(0)}(\xi)H_{n'}^{(0)}(\xi)e^{-\xi^2} =
    \delta_{nn'}.
   \ee
When convenient, we also use the standard bra-ket notation:
   \be\label{eq548}
   |n_xn_yn_z,s_z\rangle \equiv \psi_{n_xn_yn_z,s_z}(\bbox{r}\sigma) .
   \ee

\subsection{Simplex basis}
\label{sec3b}

The $y$-simplex symmetry operator (\ref{eq527}) transforms the HO states
(\ref{eq548}) in the following way
   \be\label{eq606}
   \hat{S}_y |n_xn_yn_z,s_z\rangle
      = (-1)^{n_y+\frac{1}{2}-s_z}|n_xn_yn_z,-s_z\rangle .
   \ee
Since in the present implementation
of the code \pr{HFODD}, the simplex symmetry is always assumed,
it is convenient to use the HO basis composed of
states which belong to a given simplex, i.e.,
   \bnll{eq550}
   |n_xn_yn_z,s\mbox{=$+$}i\rangle
    &=& \frac{1}{\sqrt{2}}
           \left(i^{n_y}|n_xn_yn_z,{\textstyle\frac{1}{2}}\rangle
            -i^{-n_y+1}|n_xn_yn_z,-{\textstyle\frac{1}{2}}\rangle\right),
                                           \label{eq550a} \\ \steplet
   |n_xn_yn_z,s\mbox{=$-$}i\rangle
    &=& \frac{1}{\sqrt{2}}
           \left(-i^{n_y+1}|n_xn_yn_z,{\textstyle\frac{1}{2}}\rangle
                 +i^{-n_y}|n_xn_yn_z,-{\textstyle\frac{1}{2}}\rangle\right),
                                           \label{eq550b}
   \enll
for which
   \be\label{eq551}
   \hat{S}_y |n_xn_yn_z,s\mbox{=$\pm$}{i}\rangle
     = (\pm{i})|n_xn_yn_z,s\mbox{=$\pm$}{i}\rangle.
   \ee

Since the HO wave functions are real,
the time-reversal operator (\ref{eq530}) transforms them
in the following way
   \be\label{eq552}
   \hat{T} |n_xn_yn_z,s_z\rangle
      = (-1)^{\frac{1}{2}-s_z}|n_xn_yn_z,-s_z\rangle
   \ee
The relative phases of states (\ref{eq550a}) and (\ref{eq550b})
have been chosen in such a way that the time reversal simply
flips the simplex:
   \be\label{eq553}
   \hat{T} |n_xn_yn_z,s\mbox{=$\pm$}{i}\rangle
      = \pm|n_xn_yn_z,s\mbox{=}\mp{i}\rangle.
   \ee

Having the relative phases established, we may still arbitrarily chose
the absolute phases of, say, the $s$=$+i$ simplex eigenstates.
The choice in Eq.{\spc}(\ref{eq550a}) is made by considering
the antiunitary operator $\hat{K}$
   \be\label{eq531}
   \hat{K} = \hat{T}i\sigma_z \quad,\quad \hat{K}^2 = 1.
   \ee
This operator does not act on the space coordinates and therefore
conserves quantum numbers $n_xn_yn_z$. Since it is an antilinear
operator with the square equal to one, the phases in the spin space
can always be chosen \cite{Mes69} in such a way that all the basis
states are its eigenstates
with the eigenvalues being equal to 1.
Since $\hat{K}$ commutes with $\hat{T}$, such a choice of phase
convention made in (\ref{eq550a}) applies in fact to both simplexes, i.e.,
   \be\label{eq532}
   \hat{K}|n_xn_yn_z,s\mbox{=$\pm$}{i}\rangle
       =  |n_xn_yn_z,s\mbox{=$\pm$}{i}\rangle.
   \ee
One should stress that $\hat{K}$ is not a conserved symmetry,
and therefore the HF single-particle states do not have any
particular symmetry with respect to this operator.

\subsection{Matrix elements in the simplex basis}
\label{sec3c}

The properties of the operator $\hat{K}$ and the resulting phase properties
allow to simplify the form the single-particle potentials
(\ref{eq209}) when evaluated between the simplex eigenstates.
Indeed, for such a choice
various terms in these potentials have matrix elements which are either
real or imaginary, but not complex. As can be easily checked,
in the {\em time-even} potentials
the spin-independent terms and the terms proportional to $\sigma_z$
are real, while the terms proportional to $\sigma_x$ and $\sigma_y$
are imaginary, whereas,
in the {\em time-odd} potentials
the spin-independent terms and the terms proportional to $\sigma_z$
are imaginary, while the terms proportional to $\sigma_x$ and $\sigma_y$
are real. As already mentioned, the operator $\hat{K}$ is not
a conserved symmetry of the Routhians, the HF wave functions
do not have any specific $\hat{K}$ symmetry and are complex.

Let $\hat{O}$ denote an arbitrary operator acting only in
the space coordinates (independent of spin) and let
$|n_xn_yn_z\rangle$ denote pure HO states (no spin component). Then
we may summarize the properties of matrix elements
in the simplex basis in the following way:
   \bnll{eq554}
   \langle{}n_xn_yn_z,s\mbox{=$+$}i|\hat{O}\sigma_0
           |n_x'n_y'n_z',s\mbox{=$+$}i\rangle &=&
   \langle{}n_xn_yn_z|\hat{O}|n_x'n_y'n_z'\rangle F(n_y-n_y'),
                                                   \\ \steplet
   \langle{}n_xn_yn_z,s\mbox{=$+$}i|\hat{O}\sigma_x
           |n_x'n_y'n_z',s\mbox{=$+$}i\rangle &=&
   \langle{}n_xn_yn_z|\hat{O}|n_x'n_y'n_z'\rangle F(n_y+n_y'+1),
                                                   \\ \steplet
   \langle{}n_xn_yn_z,s\mbox{=$+$}i|\hat{O}\sigma_y
           |n_x'n_y'n_z',s\mbox{=$+$}i\rangle &=&
   \langle{}n_xn_yn_z|\hat{O}|n_x'n_y'n_z'\rangle F(n_y+n_y'+2),
                                                   \\ \steplet
   \langle{}n_xn_yn_z,s\mbox{=$+$}i|\hat{O}\sigma_z
           |n_x'n_y'n_z',s\mbox{=$+$}i\rangle &=&
   \langle{}n_xn_yn_z|\hat{O}|n_x'n_y'n_z'\rangle i F(n_y-n_y'+1),
   \enll
where the factors $F(n)$ are defined as
   \be\label{eq555}
   F(n) = \left\{\ba{rcl} (-1)^{\frac{n}{2}}&\mbox{for}&n{\spc}\mbox{even,} \\
                                          0 &\mbox{for}&n{\spc}\mbox{odd.}
                 \ea\right.
   \ee
These properties allow omitting the calculation of matrix elements
for which $F$=0. As mentioned in Sec.{\spc}\ref{sec9b}, the
matrix elements corresponding to $s$=$-$$i$ need not be
explicitly calculated.

\subsection{Matrix elements in the coordinate space}
\label{sec3d}

For the Skyrme interaction, the Routhian operators (\ref{eq515}) are
differential operators composed of terms up to the second-order
derivatives. Therefore, the spatial matrix elements of
the type $\langle{}n_xn_yn_z|\hat{O}|n_x'n_y'n_z'\rangle$
can be calculated by using differential formulas for the
HO wave functions (\ref{eq546}), i.e.,
   \bnll{eq556}
   \frac{d}{dx_\mu}\psi_{n_\mu}(x_\mu)
           &=& b_\mu^{\frac{3}{2}}H_{n_\mu}^{(1)}(\xi_\mu)
                           e^{-\frac{1}{2}\xi_\mu^2},
                                                      \\ \steplet
   \frac{d^2}{dx_\mu^2}\psi_{n_\mu}(x_\mu)
           &=& b_\mu^{\frac{5}{2}}H_{n_\mu}^{(2)}(\xi_\mu)
                           e^{-\frac{1}{2}\xi_\mu^2},
   \enll
where $H_{n}^{(1)}(\xi)$ and $H_{n}^{(2)}(\xi)$ are the ($n$+1)- and
($n$+2)-order polynomials of $\xi$ defined by
   \bnll{eq558}
    H_{n}^{(1)}(\xi) &=& 2nH_{n-1}^{(0)}(\xi)
                            - \xi H_{n}^{(0)}(\xi) ,
                                                      \\ \steplet
    H_{n}^{(2)}(\xi) &=& (\xi^2-2n-1) H_{n}^{(0)}(\xi) .
   \enll
Matrix elements of all types of differential operators
(up to the second order)
can therefore be expressed through the expansion coefficients
$C^k_{nn'}(dd')$,
   \be\label{eq559}
    H_{n}^{(d)}(\xi)H_{n'}^{(d')}(\xi) = \sum_{k=0}^{n+n'+d+d'}
             C^k_{nn'}(dd')H_{k}^{(0)}(\xi) ,
   \ee

with 0$\leq$$d$+$d'$$\leq$2, and through the integrals involving
the polynomials $H_{k}^{(0)}(\xi)$ only.  Coefficients
$C^k_{nn'}(dd')$ can be in principle calculated recursively or
explicitly \cite{Gir83}, however, a method which is simpler and
less prone to programming errors has been used in the code
\pr{HFODD}, namely, they are calculated numerically with the
machine precision by employing the orthogonality relations
(\ref{eq547}) of Hermite polynomials and the Gauss-Hermite
quadratures.  {}From now on, we consider explicitly only those
terms which do not contain derivatives, while all derivative
terms can be treated analogously by using conditions
(\ref{eq559}) for $d$,$d'$$>$0.

Concentrating on terms without derivatives,
the matrix elements of an arbitrary function
$O(x,y,z)$ can be calculated as
   \be\label{eq560}
    \langle{}n_xn_yn_z|\hat{O}|n_x'n_y'n_z'\rangle =
    \sum_{k_x} C^{k_x}_{n_xn_x'}(00)
    \sum_{k_y} C^{k_y}_{n_yn_y'}(00)
    \sum_{k_z} C^{k_z}_{n_zn_z'}(00) O_{k_xk_yk_z} ,
   \ee
where
   \be\label{eq561}
       O_{k_xk_yk_z}  = \int d\xi_xd\xi_yd\xi_z
                          {\textstyle{O\left(\frac{\xi_x}{b_x},
                                             \frac{\xi_y}{b_y},
                                             \frac{\xi_z}{b_z}\right)}}
                          H_{k_x}^{(0)}(\xi_x)
                          H_{k_y}^{(0)}(\xi_y)
                          H_{k_z}^{(0)}(\xi_z)
                          e^{-\xi_x^2-\xi_y^2-\xi_y^2} .
   \ee
Integrals (\ref{eq561}) are calculated by the standard
Gauss-Hermite quadratures \cite{Abr70}. These integrals can be
calculated {\em exactly} by noticing that
for the Skyrme interaction
the functions $O(x,y,z)$ are linear combinations
[see Eqs.{\spc}(\ref{eq210})] of densities (\ref{eq518}),
which in turn are quadratic in the HO wave functions
and their derivatives. Therefore, all\footnote
{Except from the terms resulting from the density dependence
(\ref{eq211}),
and from the exchange Coulomb terms (\ref{eq506}).
The direct Coulomb terms are treated separately in Sec.{\spc}\ref{sec4b}.}
integrated functions $O(x,y,z)$
have a form of polynomials $W(x,y,z)$  multiplied
by the typical HO Gaussian factors, i.e.,
   \be\label{eq562}
                          {\textstyle{O\left(x,y,z\right)}}
                        = {\textstyle{W\left(x,y,z\right)}}
                         e^{-\xi_x^2-\xi_y^2-\xi_y^2} ,
   \ee
for $\xi_\mu$=$b_\mu{}x_\mu$, cf. Eq.\,(\ref{eq549}).
By including the Gaussian factors explicitly in the
Gauss-Hermite quadratures one can calculate the integrals
of the remaining polynomials exactly, i.e.,
   \bn
       O_{k_xk_yk_z} &=&\int d\xi_xd\xi_yd\xi_z
               {\textstyle{W\left(\frac{\xi_x}{b_x},
                                   \frac{\xi_y}{b_y},
                                   \frac{\xi_z}{b_z}\right)}}
                H_{k_x}^{(0)}(\xi_x)
                H_{k_y}^{(0)}(\xi_y)
                H_{k_z}^{(0)}(\xi_z)
                e^{-2\xi_x^2-2\xi_y^2-2\xi_y^2}   \nonumber \\
           &=&{\textstyle{\frac{1}{\sqrt{8}}}}
                \int d\eta_xd\eta_yd\eta_z
                {\textstyle{W\left(\frac{\eta_x}{\sqrt{2}b_x},
                                   \frac{\eta_y}{\sqrt{2}b_y},
                                   \frac{\eta_z}{\sqrt{2}b_z}\right)}
                H_{k_x}^{(0)}(\frac{\eta_x}{\sqrt{2}})
                H_{k_y}^{(0)}(\frac{\eta_y}{\sqrt{2}})
                H_{k_z}^{(0)}(\frac{\eta_z}{\sqrt{2}})}
                e^{-\eta_x^2-\eta_y^2-\eta_y^2}   \nonumber \\
           &=&
         \sum_{l_x=1}^{L_x}  G_{k_x}^{l_x}
         \sum_{l_y=1}^{L_y}  G_{k_y}^{l_y}
         \sum_{l_z=1}^{L_z}  G_{k_z}^{l_z}
                {\textstyle{W\left(\frac{\eta_{l_x}}{\sqrt{2}b_x},
                                   \frac{\eta_{l_y}}{\sqrt{2}b_y},
                                   \frac{\eta_{l_z}}{\sqrt{2}b_z}\right)}}.
                                                      \label{eq563}
   \en
In the final formula, $\eta_l$ are the standard Gauss-Hermite
nodes \cite{Abr70} and the corresponding weights $w_l$ are
included in the matrices $G_k^l$, i.e.,
   \be\label{eq564}
   G_k^l = {\textstyle{\frac{w_l}{\sqrt{2}}}}
            H_k^{(0)}{\textstyle(\frac{\eta_l}{\sqrt{2}})} .
   \ee

The orders $L_x$, $L_y$, and $L_z$ of the Gauss-Hermite quadratures
can be estimated from the maximum numbers of HO quanta included
in the basis, Eq.{\spc}(\ref{eq544}).
These quadratures give exact results, also for
terms depending on derivatives, for
   \be\label{eq565}
    L_\mu = 2N_\mu +2 ,
   \ee
or larger.

It turns out that the CPU time required to perform the summations
in Eq.{\spc}(\ref{eq563}) is very small compared to other parts
of the code.  It is so provided the order of summations is as
indicated in the formula, whence at most four-fold nested loops
are required.  A much larger effort is required to perform the
summations in Eq.{\spc}(\ref{eq560}), where, even with the
indicated order of sums, the seven-fold nested loops are
necessary.  Even if, in practice, the dimensions of these loops
are rather small, this requires a substantial numerical effort.
This is also the part of the code which is least amenable to
vectorization.  Such considerations indicate that the use of the
Gauss-Hermite orders (\ref{eq565}) ensuring exact integration is
fully justified, even if these can be quite large for large
numbers of HO quanta, and are certainly much larger than those
usually used by other authors and codes.

{}From the presented analysis it is clear that the
summations of the HF densities (\ref{eq518}) have to be performed
only at the spatial points which are defined by the
Gauss-Hermite integration. Moreover, the same derivatives
of HO wave functions as in Eqs.{\spc}(\ref{eq556}) can
be used to calculate the derivative densities in Eq.{\spc}(\ref{eq518}).
The spatial
symmetries discussed in Sec.{\spc}\ref{sec9} are employed
to reduce the number of points in space where the densities
have to be summed up, i.e., for three, two, or one symmetry
planes only 1/8, 1/4, or, 1/2 points, respectively, are
used. Table \ref{tab2} summarizes the symmetries of different
HF densities with respect to the $y$-$z$, $x$-$z$, or $x$-$y$
plane \cite{DDRW}. These three symmetry planes are independent
of one another and appear for the conserved symmetries
$\hat{S}_x^T$, $\hat{S}_y$, and $\hat{S}_z^T$, respectively,
although, as discussed above, not leading to the conserved
quantum numbers (except if the parity is implied).


\section{Direct Coulomb potential}
\label{sec4b}

Direct Coulomb potential is given by three-dimensional
integral (\ref{eq513}) and therefore differs from
all the Skyrme mean-field potentials (\ref{eq210})
which are simple linear combinations of densities.
Various methods of calculating this potential are used
in the existing HF codes. A direct integration of the
Poisson equation can be performed in the one-dimensional
calculations restricted to the spherical symmetry.
In two-dimensional calculations performed for axially
deformed nuclei such an integration is also possible \cite{Vau73}.
In three dimensions one may either solve the Poisson equation
by using the conjugate gradient method \cite{PH} or replace
the Coulomb interaction by an integral over Gaussian interactions
\cite{Gir83,Dob96}. The latter method is very convenient
for the Gogny interaction for which the matrix elements of
Gaussians have to be anyhow calculated.

In the code \pr{HFODD} we have implemented the method based
on using the Coulomb Green function \cite{Jac62}, which is particularly
well suited for calculations employing the Cartesian HO basis.
This is so because the Coulomb potential (\ref{eq513}) is then, in fact,
not explicitly required; it is enough to calculate its HO-basis
matrix elements.

The direct Coulomb potential (\ref{eq513}) can be expressed through
the Dirichlet Green function $G_D(\bbox{r},\bbox{r}')$
in the following way \cite{Jac62}
   \be\label{eq566}
   U^{\text{Coul}}(\bbox{r})
        =  e^2 \int_V d^3\bbox{r}'
                       G_D(\bbox{r},\bbox{r}') \rho_p(\bbox{r}')
        - \frac{1}{4\pi} \oint_S d^2s'
            \frac{\partial{}G_D(\bbox{r},\bbox{r}')}{\partial{}n'}
                       U^{\text{Coul}}(\bbox{r}').
   \ee
The first term is the volume integral over an arbitrary closed volume,
while the second integral is performed over the surface of this volume.
The Coulomb potential has to be known on the surface.
The Dirichlet Green function fulfills the Poisson equation for
a point charge and vanishes at the surface.
In the surface term, the normal derivative
is calculated with respect to the outward direction
perpendicular to the surface.

In the present application it is
convenient to use in (\ref{eq566}) the volume in the form of the
parallelepiped
    \be\label{eq567}\ba{rcl}
    -D_x \leq &x& \leq D_x , \\
    -D_y \leq &y& \leq D_y , \\
    -D_z \leq &z& \leq D_z . \\
    \ea\ee
Then, the required Green function can be expressed in a separable
form as
    \be\label{eq568}
    G_D(\bbox{r},\bbox{r}') = \frac{4\pi}{D_xD_yD_z} \sum_{j_xj_yj_z}
    \frac{f(J_xx)f(J_yy)f(J_zz)f(J_xx')f(J_yy')f(J_zz')}
         {J_x^2 + J_y^2 + J_z^2} ,
    \ee
where the functions $f$ equal to sine or cosine
ensuring the Dirichlet boundary conditions:
    \be\label{eq569}
    f(J_\mu{}x_\mu) =
    \left\{\ba{rcl} \cos(J_\mu{}x_\mu) &\mbox{for}&j_\mu{\spc}\mbox{even,} \\
                    \sin(J_\mu{}x_\mu) &\mbox{for}&j_\mu{\spc}\mbox{odd,}
                 \ea\right.
    \ee
and
    \be\label{eq569a}
        J_\mu = \frac{(j_\mu+1)\pi}{2D_\mu} .
    \ee

In expression (\ref{eq568}), the sums over $j_x$, $j_y$, and $j_z$
are performed
over all non-negative integer values. In practice, these sums have
to be restricted to finite ranges. In the code \pr{HFODD} we
use the summation bounds
   \be\label{eq579}
   0 \leq j_x, j_y, j_z \leq N^{\text{Coul}} ,
   \ee
and $N^{\text{Coul}}$ can be chosen
based on the following considerations:
It is clear that when using the
Green function for the calculation of the Coulomb potential one must
effectively perform the Fourier transform of the proton density.
Therefore, the corresponding wave vectors $J_\mu$ have to cover
the region of momenta for which the proton momentum density distribution
is large. This distribution is localized in the momentum space,
i.e., does not extend to very large momenta; therefore, the sums
can be terminated for indices $j_\mu$ giving suitable large fixed values
of $J_\mu$. The value of $N^{\text{Coul}}$ depends, therefore,
linearly on
the steps in the momentum space given by $\pi/(2D_\mu)$,
and hence inversely on the dimensions $D_\mu$. Therefore,
we are interested in working with appropriately small sizes $D_\mu$
of the parallelepiped.

On the other hand, one has also to know the values of the
Coulomb potential on the six faces of the parallelepiped. Suppose these
faces are located far enough, where the proton density is already
very small. Then the Coulomb potential on these faces can be
very well approximated by the standard multipole expansion
\cite{Jac62}
   \be\label{eq570}
    U^{\text{Coul}}(\bbox{r}) = \sum_{\lambda\mu}
    \frac{4\pi{}e^2}{(2\lambda+1)r^{2\lambda+1}} Q^p_{\lambda\mu}
             r^\lambda{}Y_{\lambda\mu}(\theta,\phi) .
   \ee
For the validity and precision of the multipole expansion
truncated to a few lowest values of $\lambda$, we are interested
in working with appropriately large sizes $D_\mu$
of the parallelepiped.

In practice it turns out that one can chose suitable dimensions
of the parallelepiped so as to fulfill both requirements simultaneously.
In the code \pr{HFODD} the dimensions in three directions
are chosen to be proportional to the corresponding
oscillator lengths 1/$b_\mu$, i.e.,
    \be\label{eq572}
    D_\mu = \frac{d}{\sqrt{2}b_\mu} ,
    \ee
where $d$ is a dimensionless parameter defining the overall size
of the parallelepiped.  Based on the performed numerical tests, the
recommended values are $d$=20 and $N^{\text{Coul}}$=80, which for
a heavy nucleus corresponds to the parallelepiped of about
$D_\mu$$\sim$30\,fm, and to the maximum momentum and the
integration step in the Fourier transforms of about 4\,fm$^{-1}$
and 0.05\,fm$^{-1}$, respectively.  For the SD states in {\Dy}
the maximum contributions of the monopole, quadrupole, and
hexadecapole terms to the Coulomb potential are at the surface of
the $D_\mu$$\sim$30\,fm cube equal to about 3\,MeV, 50\,keV,
and 1.5\,keV, respectively.  In the multipole expansion of
Eq.{\spc}(\ref{eq570}) the multipoles above $\lambda$=4 can
therefore by safely neglected.

The calculation of the HO matrix elements of the Coulomb potential
can now be very easily completed. First, we insert the multipole
expansion (\ref{eq570}) in the surface term in Eq.{\spc}(\ref{eq566}).
All surface integrals can now be performed, because they amount
to suitable Fourier transforms of the solid harmonics
$r^\lambda{}Y_{\lambda\mu}$ on the six faces of the parallelepiped.
The values of proton multipole moments
$Q^p_{\lambda\mu} $ are anyhow routinely calculated
for the determination of nuclear deformations.

Second, we use the fact that the proton density has the form
of a polynomial multiplied by the HO Gaussian factor
[cf.{\spc}Eq.{\spc}(\ref{eq562})],
   \be\label{eq562a}
                         {\textstyle{\rho^p\left(x,y,z\right)}}
                        = {\textstyle{W_\rho^p\left(x,y,z\right)}}
                         e^{-\xi_x^2-\xi_y^2-\xi_y^2} ,
   \ee
and this polynomial can be
expressed {\em exactly} as a linear combination of Hermite polynomials,
    \be\label{eq571}
    W_\rho^p(x,y,z) = \sum_{k_xk_yk_z} {\rho'}^p_{k_xk_yk_z}
                    H_{k_x}^{(0)}(\sqrt{2}b_xx)
                    H_{k_y}^{(0)}(\sqrt{2}b_yy)
                    H_{k_z}^{(0)}(\sqrt{2}b_zz) .
   \ee
Here, for later convenience we have included the factors $\sqrt{2}$
in the arguments of the normalized Hermite polynomials.

Coefficients ${\rho'}^p_{k_xk_yk_z}$ can be calculated
(again exactly) by using the Gauss-Hermite quadratures
of the order given by Eq.\,(\ref{eq565}) in the formula analogous to 
(\ref{eq561}).
In fact, if we want to use the same Gauss-Hermite nodes as before,
we have to employ the formula
   \be\label{563a}
       {\rho'}^p_{k_xk_yk_z} =
         \sum_{l_x=1}^{L_x}  {G'}_{k_x}^{l_x}
         \sum_{l_y=1}^{L_y}  {G'}_{k_y}^{l_y}
         \sum_{l_z=1}^{L_z}  {G'}_{k_z}^{l_z}
            {\textstyle{W_\rho^p\left(\frac{\eta_{l_x}}{\sqrt{2}b_x},
                                      \frac{\eta_{l_y}}{\sqrt{2}b_y},
                                      \frac{\eta_{l_z}}{\sqrt{2}b_z}\right)}},
   \ee
for
   \be\label{eq564a}
   {G'}_k^l = w_lH_k^{(0)}(\eta_l) .
   \ee

Third, we insert the proton density in the form of
Eqs.{\spc}(\ref{eq562a}) and (\ref{eq571})
in the volume integral
in Eq.{\spc}(\ref{eq566}). Then we see that the Fourier transform
of proton density
amounts to calculating the Fourier transforms of the HO oscillator
wave functions. In practice, the dimensions of the parallelepiped
are large enough to replace these transforms by the analytic
results pertaining to integrations extended to infinity, i.e.,
   \bn\label{eq573}
    B^j_k &=& \int_{-d}^d  d\eta f(\omega_j\eta)
                      H_k^{(0)}(\eta)e^{-\frac{1}{2}\eta^2} \nonumber \\[1em]
          &\simeq& \left\{\ba{ll}
             \sqrt{2\pi} H_k^{(0)}(\omega_j)e^{-\frac{1}{2}\omega_j^2}
              (-1)^{\left[\frac{k}{2}\right]}
                            &\mbox{for} \quad{k}+{j}\quad \mbox{even,} \\
                          0 &\mbox{for} \quad{k}+{j}\quad \mbox{odd,}  \\
                    \ea\right.
   \en
where
   \be\label{eq574}
    \omega_j = \frac{(j+1)\pi}{2d},
   \ee
and $\left[\frac{k}{2}\right]$ denotes the integer part.
Due to the choice of parallelepiped dimensions
adapted to the basis, Eq.{\spc}(\ref{eq572}), the factors $B^j_k$
are the same for each Cartesian direction.

After these manipulations, the Coulomb potential,
the volume and surface terms alike, are expressed
as linear combinations of terms proportional to
$f(J_xx)f(J_yy)f(J_zz)$. Instead of performing this sums,
and calculating the Coulomb potential in spatial coordinates,
one may rather directly calculate its HO-basis matrix elements.
This gives
   \be\label{eq575}
   {U'}^{\text{Coul}}_{k_xk_yk_z} = \frac{8e^2}{\pi{}d}
                    \sum_{j_x}B^{j_x}_{k_x}
                    \sum_{j_y}B^{j_y}_{k_y}
                    \sum_{j_z}B^{j_z}_{k_z}
                    \frac{V_{j_xj_yj_z}+\sum_{\lambda\mu}Q^p_{\lambda\mu}
                             S^{\lambda\mu}_{j_xj_yj_z}}
                         {b_x^2(j_x+1)^2+
                          b_y^2(j_y+1)^2+
                          b_z^2(j_z+1)^2} ,
   \ee
where the volume term reads
   \be\label{eq576}
   V_{j_xj_yj_z} =
                    \sum_{k_x}B^{j_x}_{k_x}
                    \sum_{k_y}B^{j_y}_{k_y}
                    \sum_{k_z}B^{j_z}_{k_z}
                    {\rho'}^p_{k_xk_yk_z} ,
   \ee
and the sums in Eq.{\spc}(\ref{eq575}) are restricted to
the finite set of indices (\ref{eq579}).

In Eq.{\spc}(\ref{eq575})
the surface term is a linear combination of proton multipole
moments multiplied by fixed matrices $S^{\lambda\mu}_{j_xj_yj_z}$.
These matrices are calculated only once by
integrating solid harmonics on the faces of
the parallelepiped. Finally, the Coulomb matrix elements are obtained
from
   \be\label{eq578}
    \langle{}n_xn_yn_z|\hat{U}^{\text{Coul}}|n_x'n_y'n_z'\rangle =
    \sum_{k_x} {C'}^{k_x}_{n_xn_x'}
    \sum_{k_y} {C'}^{k_y}_{n_yn_y'}
    \sum_{k_z} {C'}^{k_z}_{n_zn_z'}
      {U'}^{\text{Coul}}_{k_xk_yk_z} ,
   \ee
where the new set of coefficients ${C'}^k_{nn'}$ is required
to accommodate the additional factor of $\sqrt{2}$, i.e.,
   \be\label{eq577}
    H_{n}^{(0)}(\xi)H_{n'}^{(0)}(\xi) = \sum_{k=0}^{n+n'}
             {C'}^k_{nn'}H_{k}^{(0)}(\sqrt{2}\xi) .
   \ee
Similarly as for other terms of Routhians, evaluation of sums
in Eqs.{\spc}(\ref{eq575}) and (\ref{eq576}) requires much
less numerical effort than of those in Eq.{\spc}(\ref{eq578}).

The Green function method allows calculating the matrix
elements of the Coulomb field directly from the matrix elements
of the proton density. The summations involved require the numerical
effort which is typical for other terms of the Skyrme interaction,
in spite of the fact that the Coulomb interaction is not of zero-order.
Performing the sums in the order indicated in Eqs.{\spc}(\ref{eq575})
and (\ref{eq576}) never leads to more than
four-fold nested loops over low-dimension indices.



\section{Acknowledgments}
\label{sec8}

\bigskip
Useful comments by J.-F. Berger, C. Chinn, L. Egido, P.-H. Heenen, and
P. Magierski are
gratefully acknowledged. We would like to express our thanks
to the {\it Institut du D\'eveloppement et de Ressources en
Informatique Scientifique} (IDRIS) of CNRS, France, which
provided us with the computing facilities under Project
No.~960333.  This research was supported in part by the Polish
Committee for Scientific Research under Contract
No.~2~P03B~034~08, and by the computational grant from the
Interdisciplinary Centre for Mathematical and Computational
Modeling (ICM) of the Warsaw University.

\clearpage

\begin{table}
\caption[TT]{
Pattern of allowed non-zero values (denoted by X)
of the expectation values of the multipole moment operators for four different
cases of symmetries treated by the \pr{HFODD} code.
}
\label{tab1}
\begin{center}
\begin{tabular}{|c|c|c|c|c|}
\hline
Symmetry planes:   &  one, $x$-$z$ &  two, $x$-$z$ and $y$-$z$
                   &  two, $x$-$z$ and $x$-$y$ &  three \\
\hline
Conserved:~~~~~~~  & $\hat{S}_y$
                   & $\hat{S}_y$ and $\hat{S}_x^T$
                   & $\hat{S}_y$ and                    $\hat{S}_z^T$
                   & $\hat{S}_y$,    $\hat{S}_x^T$, and $\hat{S}_z^T$
\\
\hline
$Q_{10}$ &   X   &   X   &   0   &   0     \\
$Q_{11}$ &   X   &   0   &   X   &   0     \\
\hline
$Q_{20}$ &   X   &   X   &   X   &   X     \\
$Q_{21}$ &   X   &   0   &   0   &   0     \\
$Q_{22}$ &   X   &   X   &   X   &   X     \\
\hline
$Q_{30}$ &   X   &   X   &   0   &   0     \\
$Q_{31}$ &   X   &   0   &   X   &   0     \\
$Q_{32}$ &   X   &   X   &   0   &   0     \\
$Q_{33}$ &   X   &   0   &   X   &   0     \\
\hline
$Q_{40}$ &   X   &   X   &   X   &   X     \\
$Q_{41}$ &   X   &   0   &   0   &   0     \\
$Q_{42}$ &   X   &   X   &   X   &   X     \\
$Q_{43}$ &   X   &   0   &   0   &   0     \\
$Q_{44}$ &   X   &   X   &   X   &   X     \\
\hline
\end{tabular}
\end{center}
\end{table}

\begin{table}
\caption[TT]{
Symmetries of HF densities with respect to the three symmetry
planes appearing when one of the three different symmetries are conserved.
}
\label{tab2}
\begin{center}
\[\begin{array}{|c|ccc||c|ccc|}
\hline
\mbox{Conserved:}                 & \hat{S}_x^T & \hat{S}_y  & \hat{S}_z^T &
\mbox{Conserved:}                 & \hat{S}_x^T & \hat{S}_y  & \hat{S}_z^T \\
\hline
\mbox{Plane:}~~~~~~               & y\mbox{-}z  & x\mbox{-}z & x\mbox{-}y  &
\mbox{Plane:}~~~~~~               & y\mbox{-}z  & x\mbox{-}z & x\mbox{-}y  \\
\hline
\rho   & + & + & + & \Delta\rho                      & + & + & + \\
\tau   & + & + & + & \bbox{\nabla}\cdot\bbox{J}      & + & + & + \\ \hline
s_1    & - & - & + & \nabla_1\rho                    & - & + & + \\
s_2    & + & + & + & \nabla_2\rho                    & + & - & + \\
s_2    & + & - & - & \nabla_3\rho                    & + & + & - \\ \hline
T_1    & - & - & + & \Delta{}s_1                     & - & - & + \\
T_2    & + & + & + & \Delta{}s_2                     & + & + & + \\
T_3    & + & - & - & \Delta{}s_3                     & + & - & - \\ \hline
j_1    & + & + & - & (\bbox{\nabla}\times\bbox{s})_1 & + & + & - \\
j_2    & - & - & - & (\bbox{\nabla}\times\bbox{s})_2 & - & - & - \\
j_3    & - & + & + & (\bbox{\nabla}\times\bbox{s})_3 & - & + & + \\ \hline
J_{11} & - & - & - & (\bbox{\nabla}\times\bbox{j})_1 & - & - & + \\
J_{21} & + & + & - & (\bbox{\nabla}\times\bbox{j})_2 & + & + & + \\
J_{31} & + & - & + & (\bbox{\nabla}\times\bbox{j})_3 & + & - & - \\ \hline
J_{12} & + & + & - & \multicolumn{4}{c}{\ }  \\
J_{22} & - & - & - & \multicolumn{4}{c}{\ }  \\
J_{32} & - & + & + & \multicolumn{4}{c}{\ }  \\ \cline{1-4}
J_{13} & + & - & + & \multicolumn{4}{c}{\ }  \\
J_{23} & - & + & + & \multicolumn{4}{c}{\ }  \\
J_{33} & - & - & - & \multicolumn{4}{c}{\ }  \\ \cline{1-4}
\end{array}\]
\end{center}
\end{table}


%
%
\end{document}